\documentstyle[twoside,fleqn,espcrc2,psfig]{article}
\def\beq{\begin{equation}}
\def\eeq{\end{equation}}
\def\bea{\begin{array}}
\def\eea{\end{array}}
\def\beqa{\begin{eqnarray}}
\def\eeqa{\end{eqnarray}}

\def\u1{{U(1)}}
\def\su2{{SU(2)}}
\newcommand{\re}{\relax{\rm I\kern-.18em R}}

\newcommand{\AmS}{{\protect\the\textfont2
  A\kern-.1667em\lower.5ex\hbox{M}\kern-.125emS}}

\title{Even flavor QED3 in an external magnetic field
\vskip-3cm\hfill\small NTUA-TH-76/99\vskip2.6cm
}
\author{K.Farakos and G.Koutsoumbas\address{Physics Department, 
National Technical University,\\Zografou Campus, 157 80, Athens, Greece}, 
N.Mavromatos and A.Momen\address{Department of Physics, Theoretical
Physics,\\1 Keble Road,Oxford OX1 3NP, U.K.}}
\begin{document}
\begin{abstract}
The magnetically induced fermionic condensate
is studied at zero and finite temperatures. The effects of a
non-homogeneous external magnetic fields is briefly considered.
\end{abstract}
\maketitle
\section{INTRODUCTION}
Considerations of the effects of external magnetic fields in the Early
Universe and in problems of high $T_c$ superconductivity (\cite{dm}, \cite{fm})
are the main motivations to study the behaviour of the fermionic matter under
the influence of an external magnetic field.

The three-dimensional continuum Lagrangian of the model is given by:
\beq
{\cal L} = -\frac{1}{4}(F_{\mu\nu})^2
+ {\overline \Psi} D_\mu\gamma _\mu \Psi -m {\overline \Psi} \Psi,
\label{contmodel}
\eeq
where $D_\mu = \partial _\mu -ig a_\mu^S-i e A_\mu;$ $a_\mu^S$ is a
fluctuating gauge field, while $A_\mu$ represents the external gauge field.
The main object of interest here is the condensate $<{\overline \Psi} \Psi>,$
which is the coincidence limit of the fermion propagator, $S_F(x,y).$

\section{RESULTS IN THE CONTINUUM}

A first estimate for the enhancement of the condensate arising from the
external fields may be gained through
the analysis of the relevant Schwinger-Dyson equation:
$$
S_F^{-1}(p) = \gamma \cdot p 
$$
\beq
- g \int \frac{d^3k}{ ( 2\pi)^3}
\gamma^\mu S_F (k)
\Gamma^\nu ( k, p-k) D_{\mu \nu}(p-k)
\label{2.1}
\eeq
where $\Gamma^\nu$ is the fermion-photon vertex function and $D_{\mu
\nu}$ is the exact photon propagator (\cite{fkm2}).

The results of a recent approximate solution of the above equation in
the regime of small homogeneous external magnetic field, both for
quenched and dynamical fermions are depicted in figure 1, where one may
see the dynamical mass generated, versus the magnetic field strength.
The upper curve (labeled ``Q,Int-solve") is the solution for quenched
fermions, while the lower curve is the weak field approximation to the
dynamical fermionic condensate. We have also included the weak field
approximation to the quenched result, as a measure of the reliability
of the weak field expansion.

\begin{figure}[t]
\centerline{\psfig{figure=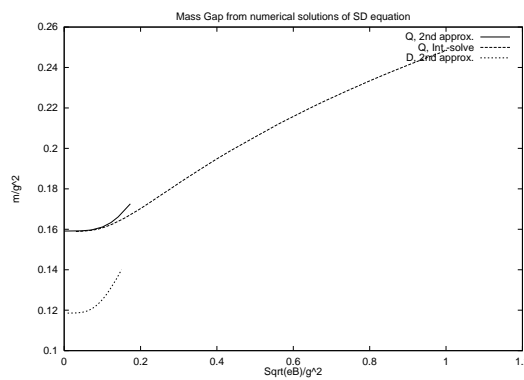,height=5cm,angle=-90}}
\caption{Solution of Schwinger-Dyson equation for quenched and
dynamical fermions. \label{f1}}
\end{figure}

There have also been approximations in the regime of strong magnetic fields
\cite{sphagin}, but for a fully
quantitative treatment one should rely on the lattice approach 
(\cite{fkm,fkm2}).

\section{LATTICE RESULTS}

We will first present the results for the $T=0$ case and a homogeneous
magnetic field. Figure 2 contains the condensate versus the gauge field
coupling constant, $\beta_g,$ for several values of the magnetic field.
This figure is the final outcome of a series of measurements performed
at several values of the bare mass; the result shown here is the 
extrapolation to the zero mass limit. 
The result is independent from the magnetic field
at strong gauge coupling, because the gauge interactions are the main
contributor to the condensate in this regime; the magnetic field takes
over in the weak coupling, on the right hand part of the diagram.

\begin{figure}[t]
\centerline{\psfig{figure=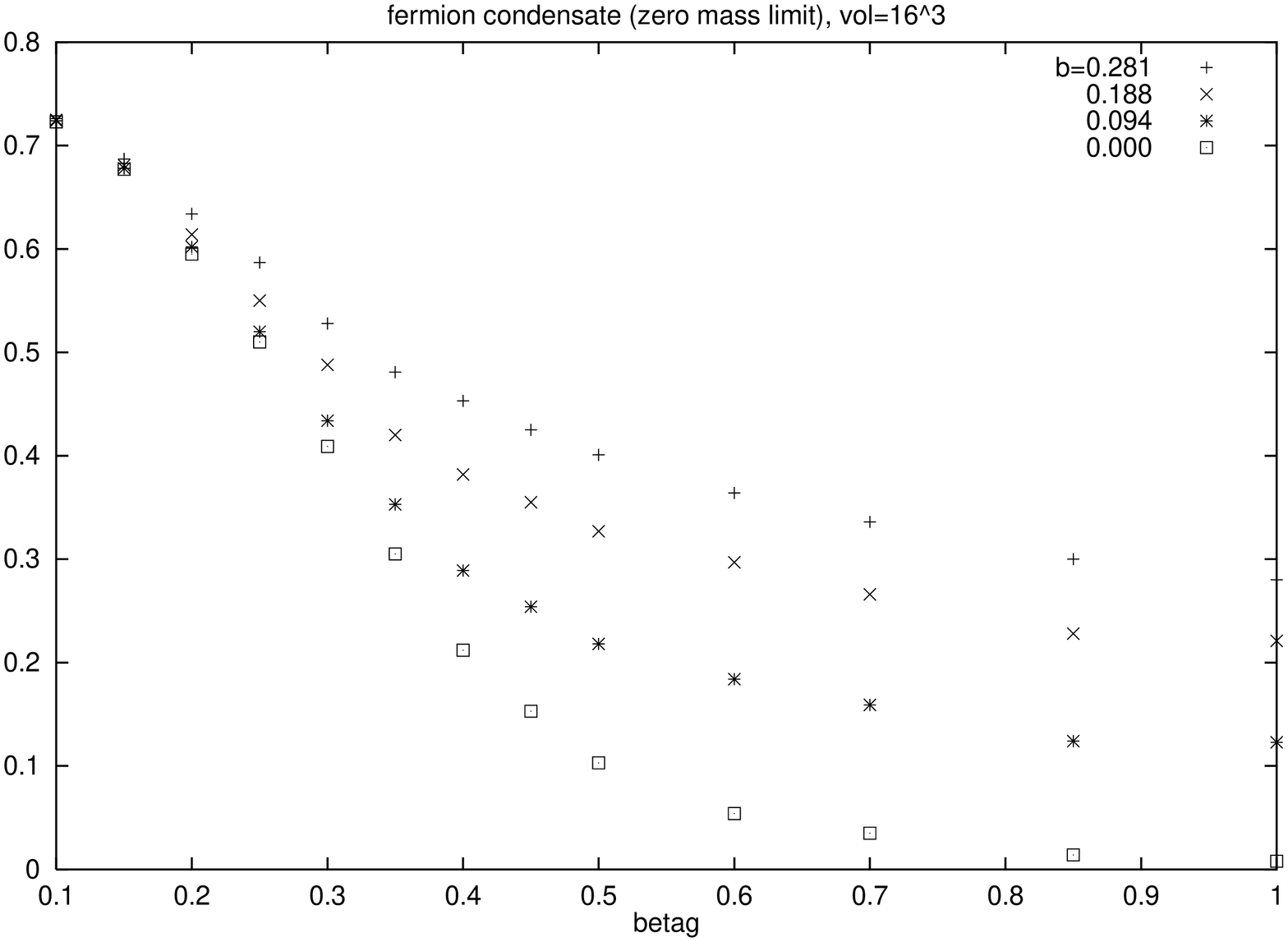,height=5cm,angle=-90}}
\caption{$<{\overline \Psi} \Psi>$ versus $\beta_g$ for
various values of the magnetic field.
\label{f2}}
\end{figure}

Figure 3 contains the condensate versus $\beta_g$ for a typical fixed value
of the magnetic field. The uppermost data correspond to a symmetric lattice
$16^3.$ The curve is smooth and no sign of discontinuity can be seen
anywhere. The next result comes from an asymmetric lattice, with a rather
large time extent, though: $24^2 \times 6.$ A structure starts showing up at
$\beta_g \approx 0.45.$ To see better this structure, we go to the
$16^2 \times 4$ lattice; the structure moves to $\beta_g \approx 0.40$
and becomes somehow more steep. The effect of the spatial size of the lattice
is not big, as one may see by comparing the points for $16^2 \times 4$ versus
the ones for $24^2 \times 4,$ which are also shown on the figure. Finally
the points for $16^2 \times 2$ show a clear discontinuity in the slope.
Although it deserves more detailed study, it seems safe to interpret this
discontinuity as a symmetry restoring phase transition, imposed by
the nonzero temperature.

\begin{figure}[t]
\centerline{\psfig{figure=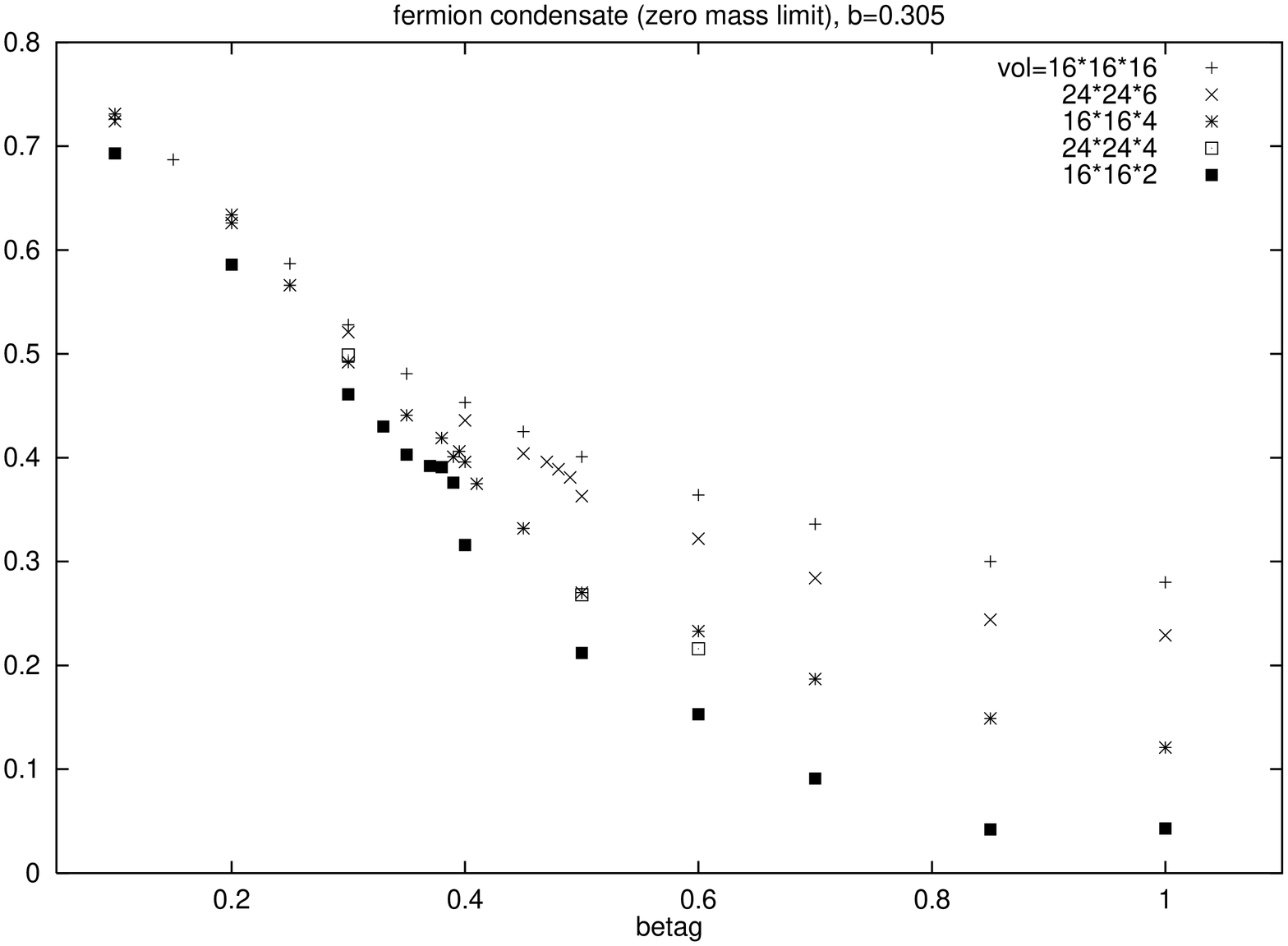,height=5cm,angle=-90}}
\caption{$<{\overline \Psi} \Psi>$ versus $\beta_G$ in the
zero mass limit at various tamperatures $\frac{1}{N_t}.$ \label{f3}}
\end{figure}
In figure 4 we use the results for various lattice sizes to construct a 
graph showing the dependence of the condensate on the temperature for
two values of the external magnetic field. It is 
surprisingly similar to the corresponding result for the ``free" 
case (\cite{fkm}). The condensate tends to zero for large temperatures; this
tendency is more intense for the smallest magnetic field.
\begin{figure}[t]
\centerline{\psfig{figure=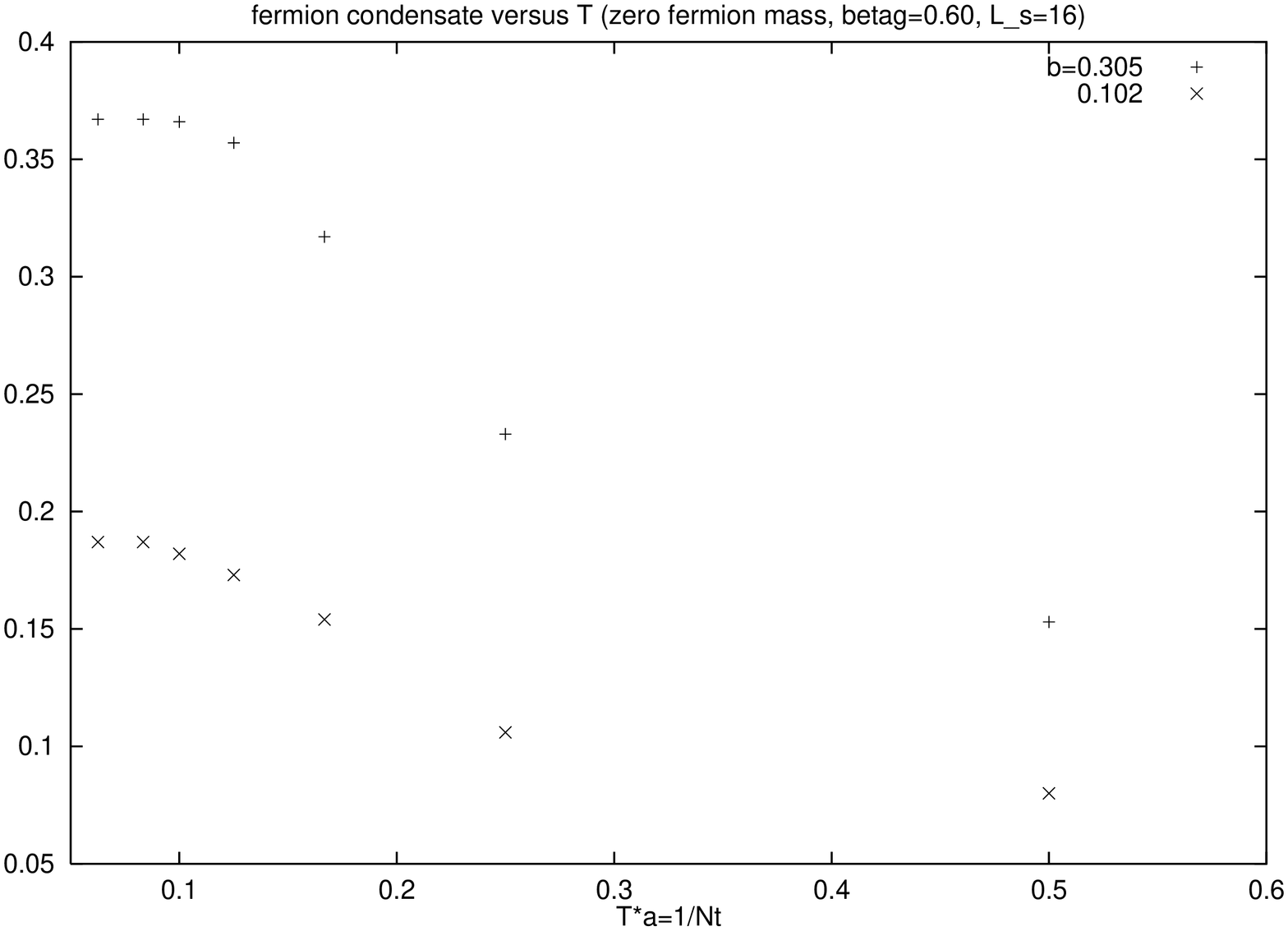,height=5cm,angle=-90}}
\caption{$<{\overline \Psi} \Psi>$ versus the temperature for
two values of the magnetic field. \label{f4}}
\end{figure}
In the last figure we show the results of a study of a non-homogeneous 
magnetic field. We consider a lattice in which its central $6 \times 6$
region (for all values of z) carries a constant magnetic flux, while
in the rest part the magnetic field vanishes (\cite{fkm2}); we measure the 
condensate along a straight line passing from the center of the lattice 
at a fixed value of the magnetic field. This profile is shown in figure 5
and one may see that the condensate is non-zero only in the region
of non-vanishing magnetic flux. Away from this region the remaining
condensate may be accounted for by the explicit mass term and has little to
do with the external magnetic field.

\begin{figure}[ht]
\centerline{\psfig{figure=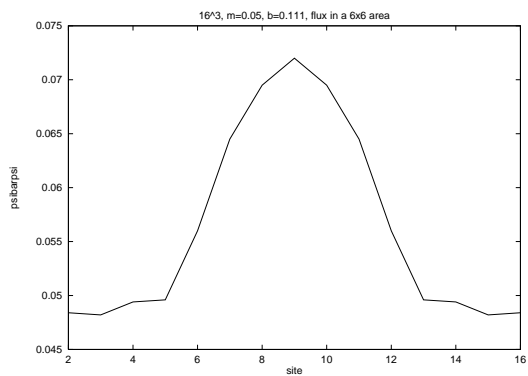,height=5cm,angle=-90}}
\caption{$<{\overline \Psi} \Psi>$ along a straight line passing
from the center of the lattice if the magnetic field parameter b 
is set to 0.111. The central region of non-zero flux is $6 \times 6.$ 
\label{f5}}
\end{figure}

\section*{Acknowledgements}

K.F. and G.K. would like to acknowledge financial support from the
TMR project ``Finite temperature phase transitions in particle Physics",
EU contract number: FMRX-CT97-0122. The work of N.E.M. is partially 
supported by PPARC (UK) through an Advanced Fellowship. The work of 
A.M. is supported by PPARC.

\end{document}